\documentclass[conference,letterpaper]{IEEEtran}
\usepackage[letterpaper, left=0.75in, right=0.7in, bottom=0.95in, top=0.75in]{geometry}

\IEEEoverridecommandlockouts
\usepackage{cite}
\usepackage{amsmath,amssymb,amsfonts}
\usepackage{algorithmic}
\usepackage{graphicx}
\usepackage{textcomp}
\usepackage{xcolor}
\usepackage{multirow}
\usepackage{epsfig}
\usepackage[super]{nth}
\usepackage{breqn}
\usepackage[linesnumbered,ruled,vlined]{algorithm2e}

\begin{document}

\title{Improved Gaussian-Bernoulli Restricted Boltzmann Machines for UAV-Ground Communication Systems}

\author{\IEEEauthorblockN{Osamah A. Abdullah\IEEEauthorrefmark{1},
		  Michael C. Batistatos\IEEEauthorrefmark{2},
		  and Hayder Al-Hraishawi\IEEEauthorrefmark{3}\\ 
				\IEEEauthorrefmark{1}\small Dept. of Electrical Engineering, Alma'moon University College, Baghdad, Iraq}
				\IEEEauthorrefmark{2}\small Dept. of Informatics and Telecommunications, University of Peloponnese, 22100 Tripolis, Greece\\
		\IEEEauthorrefmark{3}\small Interdisciplinary Centre for Security, Reliability and Trust (SnT), University of Luxembourg, L-1855, Luxembourg\\
		Email:  osamah.abdullah@wmich.edu, mbatist@uop.gr, and hayder.al-hraishawi@uni.lu 
		}

\maketitle
	\thispagestyle{plain}
   \pagestyle{plain}
    
\begin{abstract}
	Unmanned aerial vehicle (UAV) is steadily growing as a promising technology for next-generation communication systems due to their appealing features such as wide coverage with high altitude, on-demand low-cost deployment, and fast responses. UAV communications are fundamentally different from the conventional terrestrial and satellite communications owing to the high mobility and the unique channel characteristics of air-ground links. However, obtaining effective channel state information (CSI) is challenging because of the dynamic propagation environment and variable transmission delay. In this paper, a deep learning (DL)-based CSI prediction framework is proposed to address channel aging problem by extracting the most discriminative features from the UAV wireless signals. Specifically, we develop a procedure of multiple Gaussian Bernoulli restricted Boltzmann machines (GBRBM) for dimension reduction and pre-training utilization incorporated with an autoencoder-based deep neural networks (DNNs). To evaluate the proposed approach, real data measurements from an UAV communicating with base-stations within a commercial cellular network are obtained and used for training and validation. Numerical results demonstrate that the proposed method is accurate in channel acquisition for various UAV flying scenarios and outperforms the conventional DNNs.	
\end{abstract} 
	

\section{Introduction}\label{sec:intro}
Low-altitude unmanned aerial vehicles (UAVs), also commonly referred to as drones,  have enabled a plethora of personal and commercial applications including aerial photography and sightseeing, parcel delivery, emergency rescue in natural disasters, monitoring and surveillance, and precision farming \cite{Zhao2019}. Recently, the interest in this emerging technology is steadily surging  as many governments have already facilitated the regulations for UAV usage. As a result, UAV technologies are being developed and deployed at a very rapid pace around the world to offer fruitful business opportunities and new vertical markets \cite{Alladi2020}. In particular, UAVs can be employed as aerial platforms to enhance wireless connectivity for ground users and Internet of Things (IoT) devices in harsh environments when terrestrial networks are unreachable. Additionally, intelligent UAV platforms can provide important and diverse contributions to the evolution of smart cities by offering cost-efficient services ranging from environmental monitoring to traffic management \cite{Kisseleff2020}.

Wireless communication is a key enabling technology for UAVs and their integration has drawn a substantial attention in recent years. In this direction, the third generation partnership project (3GPP) has been active in identifying the requirements, technologies, and protocols for aerial communications to enable networked UAVs in current long-term evolution (LTE) and 5G/B5G networks \cite{Abdalla2021,NGSO_surevy}. UAV communications are fundamentally different from terrestrial communications in the underlying air-to-ground propagation channel and the inherent size, weight and power constraints. Apparently, the 3D mobile UAVs enjoy a higher probability of line-of-sight (LoS) communication than ground users, which can be beneficial for the reliability and power efficiency of UAV communications. Nevertheless, this also implies that UAV communications may easily cause/suffer interference to/from terrestrial networks \cite{Zeng2019}.

Realizing full-fledged UAVs in the 3D mobile cellular network depends to a large extent on the accuracy of channel state information (CSI) acquisition over diverse UAV operating environments and scenarios, which is of paramount importance to enhance system performance. Reliable CSI is crucial for aerial communications not only in control/non-payload but also in payload data transmissions, which is one of the major challenges in these systems. Moreover, obtaining precise CSI has a great significance on physical layer transmissions, radio resources allocation, and interference management, which will help in designing robust beamforming and beam tracking algorithms as well as efficient link adaptation techniques. While several statistical air-to-ground channel models that consider the trade-off between accuracy and mathematical tractability have been studied in the literature \cite{Khawaja2019}, a more practical analysis to bridge this knowledge gap is still needed. 
	
On a parallel avenue, a significant attention has been paid recently to deep learning (DL) by wireless communication community owing to its successful in wide range of applications, e.g., computer vision, natural language processing, and automatic speech recognition. DL is a neuron-based machine learning approach that is able to construct deep neural networks (DNNs) with versatile structures based on the application requirements. Specifically, several works in the open literature have utilized DL methods for channel modeling and CSI acquisition. For instance, a DL driven channel modeling algorithm is developed in \cite{Sun2021} by using a dedicated neural network based on generative adversarial networks that is designed to learn the channel transition probabilities from receiver observations. In \cite{Ma2020}, a DNN-based channel estimation scheme is proposed to jointly design the pilot signals and channel estimator for wideband massive multiple-input multiple-output (MIMO) systems.

    In this paper, we propose a DL-based UAV channel predictor by employing the advanced Gaussian–Bernoulli restricted Boltzmann machines (GBRBM) \cite{Choo2018}. The GBRBM is a useful generative stochastic model that captures meaningful features from the given multi-dimensional continuous data. It can also learn a probability distribution over a set of inputs in an unsupervised manner and it addresses the limitations of the bipartite restricted Boltzmann machine (RBM) model by replacing the binary nodes with Gaussian visible nodes that can initialize DNN for feature extraction and dimension reduction. The distinct contributions of this work can be summarized as follows:

	\begin{itemize}
		\item Applying the GBRBM model to estimate the received signal power at an UAV from a cellular network during the flight, where DNNs are employed to extract features from the UAV channels as a set of blocks for channel modeling.
		\item Developing an adaptive learning rate approach and a new enhanced gradient to improve the training performance. Specifically, an autoencoder is used to fine-tune the parameters during the training phase by using an autoencoder-based DNN.
		\item Verifying the effectiveness of the proposed framework through experimental measurements using real measurements data. The obtained results show that the proposed scheme outperforms the conventional autoencoders in realizing channel feature extraction.
	\end{itemize}
	
	The remainder of the paper is organized as follows. Section II presents the system model and problem formulation along with describing the developed approach. Simulation results and demonstrations are given in Section III. Finally, conclusion remarks are drawn in Section IV.
	
\section{System Model and Problem Formulation} \label{Sec:system_model}
	
\begin{figure}[!t]\centering 
 	\includegraphics[width=0.5\textwidth]{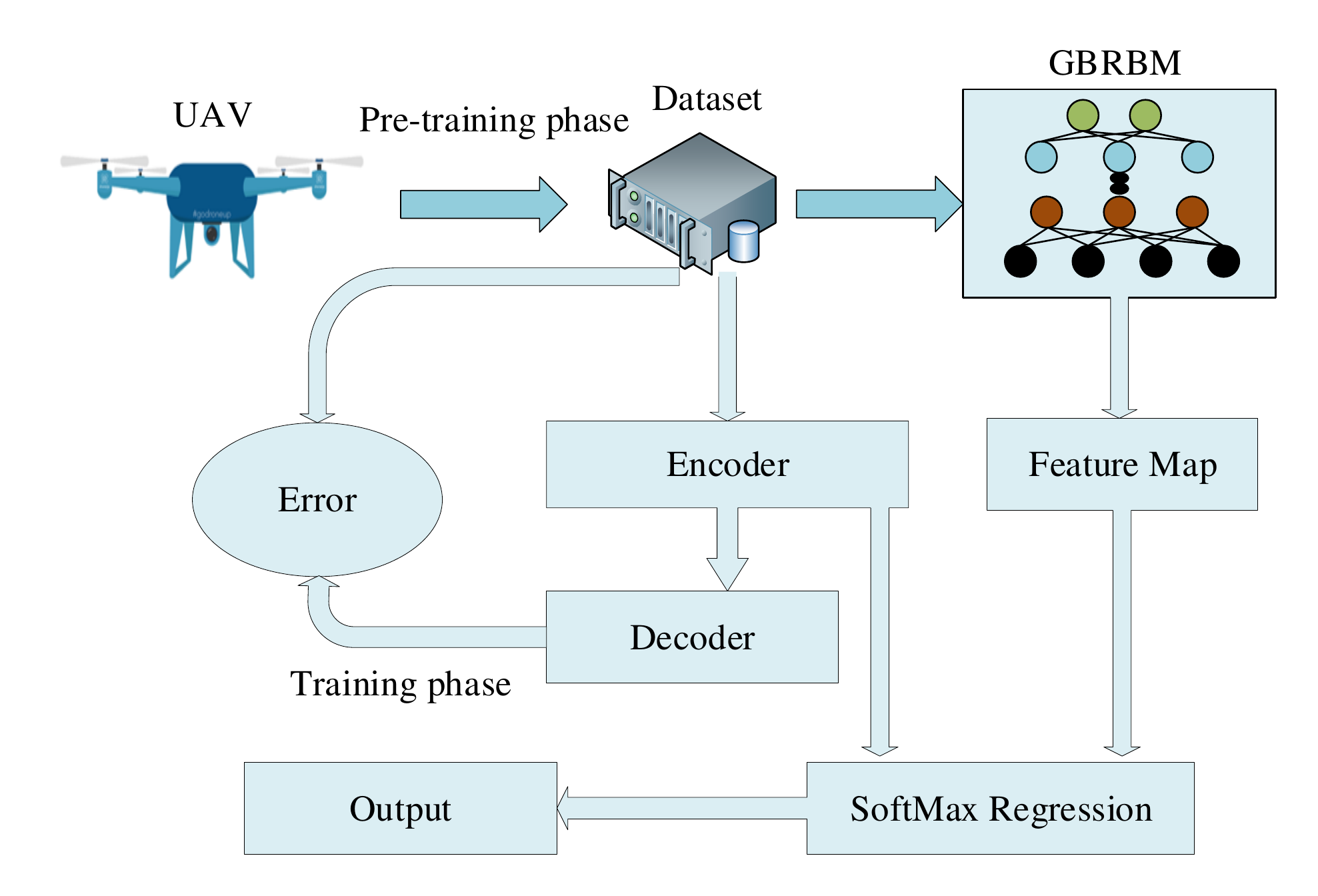}
 	\caption{Framework architecture of the proposed enhanced GBRBM.}\label{fig:Fig1}
\end{figure}

	We consider the downlink transmission of a multi-user wireless network that consists of multi-antenna base-stations (BSs) are serving multiple randomly distributed user nodes in the presence of a UAV communicating with the BSs as well. However, the channel between the BS and UAV is affected by several parameters  such as UAV altitude, antenna directivity, location, transmission power and the characteristics of the environment. To investigate the effects of these parameters on channel modeling between the UAV-transceiver and the BSs, we proposed a DL-based framework employing multiple GBRBMs for dimension reduction and pre-training utilization incorporated with an autoencoder-based deep neural network.
	The detailed framework of CSI estimation scheme is shown in Fig. \ref{fig:Fig1}, which is designed to operate systematically on the principles of the improved GBRBM model. This framework consists of two main parts: offline training and online estimation. The  offline training intensifies the framework by historical data so that it can grasp the correlation of channel variations in the different UAV flying scenarios. Thus, when a CSI estimation request arrives, the framework takes the current UAV information as input data  to predict the channel state online.
	

Among neural network models, GBRBM shows a good potential in time-series prediction owing to its ability of acquiring the unknown sequences based on historical data. GBRBM is known as Markov random fields (MRFs)  and it is an undirected probabilistic graphical model \cite{Koller2009}. 
The proposed architecture consists of an input layer, an output layer and several hidden layers for capturing channel characteristics
The input layer takes the observed data through nine different nodes $N \in \{N1,\cdots,N9\}$. Specifically,, $N1$, $N2$ and $N3$ represent the latitude, longitude, and UAV elevation angle, respectively. Additionally,  $N4$ and $N5$ account for cell latitude and longitude, respectively, while $N6$ and $N7$ represent the cell elevation and the cell building. Further, $N8$ and $N9$ are the antenna mast height the UAV altitude, respectively. Thus, the energy function of GBRBM is defined as: 
	
	\begin{small}
	\begin{eqnarray}
		E\left( v,h|\theta \right)  = \sum_{i=1}^{n_v} \frac{(v_i-b_i)^2}{2\sigma^2} - \sum_{i=1}^{n_v} \sum_{j=1}^{n_h} W_{ij} h_j \frac{v_i}{\sigma_i} - \sum_{j=1}^{n_h} c_j h_j,
	\end{eqnarray}
	\end{small}
	
	\noindent
	where $b_i$ represents the bias of the visible layer and $c_j$ represents the bias of the hidden layer, and $w_{ij}$ represents the weight that connects the visible layer to the hidden layer. Further, $\sigma_i$ accounts for the standard deviation of the visible units. Each block has nine data inputs as explained earlier. While $n > 1$ for hidden layer unit $h_j$, based on the property of MRFs and the energy function, the joint probability distribution is defined as:
	\begin{eqnarray}
		P(v,h) = \frac{1}{Z} e^{-E(v,h)},
	\end{eqnarray}
	where Z is the partitioning function, as given below
	\begin{eqnarray}
		Z = \sum_v \sum_v -E(v,h).
	\end{eqnarray}
	By using the joint probability function, the marginal distribution of $v$ can be defined as:
	\begin{eqnarray}
		P(v) =\frac{1}{Z} e^{-E(v,h)}.
	\end{eqnarray}
	The hidden and visible layers are both conditionally independent. The condition probability of $v$ and $h$ are defined as follows:
	\begin{eqnarray} 
		P(v_i=v|h) = \mathcal N \left( v|b_i + \sum_{J}^{} h_j W_{ij}  \sigma^2_i \right), \label{eq5}
	\end{eqnarray}
\begin{eqnarray}
		P(h_j=1|v) =  \text{sigmoid} \left( c_j + \sum_i W_{ij} \frac{v_i}{\sigma^2_i}\right), \label{eq6}
	\end{eqnarray}
where $\mathcal N (.|\mu, \sigma^2)$ represents the Gaussian probability density function with mean $\mu$ and variance $\sigma^2$. Further, stochastic maximization of likelihood is used to train GDBM. The likelihood is estimated by marginalizing out the hidden neurons. The partial-derivative of the maximization log-likelihood function is given by:	
\begin{eqnarray}
	\frac{\partial L}{\partial \theta } \alpha \left\langle \frac{\partial \left(-E\left(v^{(t)},h|\theta \right)  \right) }{\partial \theta}\right\rangle_d -\left\langle \frac{\partial \left(-E\left(v,h|\theta \right)  \right) }{\partial \theta}\right\rangle_m
\end{eqnarray}
where $ \left\langle . \right\rangle_d $   and  $ \left\langle . \right\rangle_m $ denote the expectation computed over the
data $P(h|\{v^{(t)}\},\theta)$ and model distributions $P(v,h|\theta)$, respectively. Here, $ \theta$ is the parameters of GBRBM because gradient calculation needs a high computational cost. Reference \cite{Carreira2005} used a contrastive-divergence (CD) learning that proved to be an efficient approximator for the loglikelihood gradient for GBRBM. The CD learning is recalled as an alternative to calculating the second term of the log-likelihood gradient by iteration a few samples from the data by using Gibbs sampling. As a result, GBRBM parameters are derived as follows:	
\begin{subequations}\label{eqn:parameter}
	\begin{eqnarray}
	&& \hspace{-12mm}	W_{ij} \xleftarrow[]{=}  W_{ij} + \eta \left( \left\langle  \frac{1}{\sigma^2_i} v_i h_j\right\rangle_d  - \left\langle  \frac{1}{\sigma^2_i} v_i h_j\right\rangle_m \right),\label{eqn:parameter_w}\\
&& \hspace{-12mm}		b_i \xleftarrow[]{=}  b_i + \eta \left( \left\langle  \frac{1}{\sigma^2_i} v_i\right\rangle_d  - \left\langle  \frac{1}{\sigma^2_i} v_i \right\rangle_m \right),\label{eqn:parameter_b}\\
&& \hspace{-12mm}		c_j \xleftarrow[]{=}  c_j + \eta \left( \left\langle  \frac{1}{\sigma^2_i} h_j\right\rangle_d  - \left\langle  \frac{1}{\sigma^2_i}  h_j\right\rangle_m \right),\label{eqn:parameter_c}
	\end{eqnarray}
\end{subequations}
where $\eta$ denotes the learning rate.  The GBRBM is updated and trained efficiently by updating (\ref{eqn:parameter_w}), (\ref{eqn:parameter_b}), and (\ref{eqn:parameter_c}).

\subsection{Adaptive Learning Rate}
Based on the maximization of the local estimate of the likelihood, the learning rate can automatically be adapted while the RBM is trained by using the stochastic gradient. Because $\theta=(W,b,c)$ is the parameter of GBRM modeling,  represents the adapted parameter of learning rate $\eta$, $P_{\theta}(V)=P^*_{\theta}/Z_{\theta}$  represents   the probability density function (pdf), and $Z_{\theta}$  is the normalization parameter for the parameter $\theta$. The optimal learning rate can be found that can maximize the likelihood of each iteration. Nevertheless, this can lead to a big fluctuation due to the small size of the minibatch.  \cite{Cho2013} proposed that the new learning rate is chosen from the set $\left\lbrace (1-\mathcal{E})^2 \eta_o, (1-\mathcal{E}) \eta_o,\eta_o, (1-\mathcal{E}) \eta_o, (1-\mathcal{E}) \eta_o, \right\rbrace $ where $\eta_o$  is the     prior learning rate and $\mathcal{E}$   is a small constant, which was chosen randomly.

\subsection{Enhanced Gradient}
The Recently enhanced gradient was proposed by \cite{Cho2013a} to update the invariant rule of the Boltzmann machines for data representation. The gradient was introduced by a bit-flipping transformation and then the rule was updated to improve the results, It has been shown that the learning of RBM can be improved by making the results less sensitive to the learning parameters and initialization. Thus, a new method to enhance the gradient is proposed in \cite{Cho2013a}  to replace the (\ref{eqn:parameter_w}), (\ref{eqn:parameter_b}), and (\ref{eqn:parameter_c}). They defined the covariance between the two variables under the distribution $P$ as given below
\begin{eqnarray}\label{eq12}
	COV_P (v_i,h_j)=\left\langle v_i h_j\right\rangle_P \left\langle v_i \right\rangle_P  \left\langle h_j\right\rangle_P 
\end{eqnarray}
The standard gradient in (\ref{eqn:parameter_w}) cab be rewritten as
\begin{eqnarray}\label{eqn:delta_w}
	\nabla_{W_{ij}}&=& COV_d (v_i,h_j) - COV_m (v_i,h_j) \nonumber\\
	 &+& \left\langle v_i \right\rangle_{dm} \nabla_{c_{j}} + \left\langle h_j \right\rangle_{dm} \nabla_{b_{i}},
\end{eqnarray}
where $ \left\langle . \right\rangle_{dm} = \frac{1}{2} \left\langle . \right\rangle_d  + \frac{1}{2} \left\langle . \right\rangle_m $  denotes the average of the model distribution and the data. The standard deviation has some potential problems. The gradient is correlated with the weights and the bias terms. Additionally,  $COV_d (v_i,h_j) - COV_m (v_i,h_j) $ is uncorrelated with $\nabla_{b_{i}}$ and $\nabla_{c_{j}}$, which may lead to distract the learning with non-useful weights when there are a lot of active neurons for which $ \left\langle . \right\rangle_{dm} \approx 1$. However, updating the weights using (\ref{eqn:delta_w}) with the obtained data may bring about some issues by flipping some of the binary units in RBM from zeros to ones and vice versa.
\begin{subequations}
	\begin{eqnarray}
	\tilde{v}_i &=&  \tilde{v}^{(1-f_i)} (1-v_i)^{f_i} \hspace{1cm} f_i \in \{0,1\} ,\\
	\tilde{h}_i &=&  \tilde{h}^{(1-g_i)} (1-h_i)^{g_j} \hspace{1cm} g_i \in \{0,1\},
	\end{eqnarray}
\end{subequations}
The parameters are transformed accordingly to $\tilde{\theta}$
\begin{subequations}
	\begin{eqnarray}
		\tilde{w}_{ij} &=& (-1)^{f_i+g_j} w_{ij},\\
		\tilde{b}_i &=& (-1)^{f_i} \left( b_i+\sum_{j}^{} g_j W_{ij} \right),
	\end{eqnarray}
\end{subequations}
The energy function is equivalent to $E(\tilde{x}+\tilde{\theta})= E({x}+{\theta}) +a$, where $a$ is a constant for all the values. That will lead to an update of the model and then transformed again. The resulting model will be defined as
\begin{subequations}
	\begin{eqnarray}
	\!\!\!\!\! w_{ij} \!\!\!\!\! &\xleftarrow[]{} & \!\!\!\! w_{ij} + \eta \left[    COV_d (v_i,h_j) - COV_m (v_i,h_j) \right. \nonumber \\ 
	 &&\left. + \left( \left\langle v_i  \right\rangle_{dm} - f_i\right) \nabla c_j + \left( \left\langle h_j \right\rangle_{dm} -g_j\right)  \nabla_{b_{i}}\right]\!,\label{eqn:parameter_wi}\\
	\!\!\!\!\! b_i \!\!\!\!\! &\xleftarrow[]{} &\!\!\!\!  b_i \!+\! \eta \left[\nabla b_i - \!\! \sum_{i}^{}g_j\left( \nabla w_{ij}- f_i \nabla c_{j} - g_i \nabla b_i \right)  \right]\label{eqn:parameter_bi}\\
\!\!\!\!\!	c_j  \!\!\!\!\! &\xleftarrow[]{} & \!\!\!\! c_j \!+\! \eta \left[\nabla c_j - \!\! \sum_{i}^{}f_i\left( \nabla w_{ij}- f_i \nabla c_{j} - g_i \nabla b_i \right)  \right]\label{eqn:parameter_ci}
	\end{eqnarray}
\end{subequations}
where $\nabla \theta$ represents the gradient parameters defined in (\ref{eqn:parameter}). Therefore, it will be $2^{nv+nh}$ different update rule, where $nv$ and $nh$ are the numbers of hidden and visible neurons. To find the maximum likelihood updates, \cite{Cho2013} proposed a new gradient weighted sum $2^{nv+nh}$ with the following weights:
\begin{eqnarray}
	\prod\limits_{i} \left\langle v_i \right\rangle_{dm}^{f_i} \left(1-\left\langle v_i \right\rangle \right)^{1-f_i} 	\prod\limits_{j} \left\langle h_j \right\rangle_{dm}^{g_j} \left(1-\left\langle h_i \right\rangle \right)^{1-g_j} 
\end{eqnarray}
Due to the larger weights, the enhanced gradient is defined as:
\begin{subequations}\label{eqn:parameter2}
	\begin{eqnarray}
&& \hspace{-12mm}	\nabla_e w_{ij} = COV_d (v_i,h_j) - COV_m (v_i,h_j), \label{eqn:parameter2_w}\\
&& \hspace{-12mm}	\nabla_e b_i  = \nabla b_i \!-\!\! \sum_{j}^{} \left\langle  h_j\right\rangle_{dm}  \!\!\bigg(\!\! \nabla W_{ij} \!-\! \nabla b_i- \left\langle v_i \right\rangle_{dm} \nabla c_j \! \bigg)\!, \label{eqn:parameter2_b}\\
&& \hspace{-12mm}	\nabla_e c_j =   \nabla c_j \!-\!\!\sum_{i}^{} \left\langle  v_i\right\rangle_{dm}  \!\! \bigg( \!\! \nabla W_{ij} \!-\! \nabla c_j- \left\langle h_j \right\rangle_{dm} \nabla c_j \!\bigg)\!. \label{eqn:parameter2_c}
	\end{eqnarray}
\end{subequations}
where $\nabla_e w_{ij}$  has the same form of (\ref{eq12}), where each block is connected to the upper block through the unit of the hidden layer to update the parameter in (\ref{eqn:parameter2}) layer by layer for GBRBM pre-training. The pre-training of GBRBM provides the initializing to deep autoencoder. The procedure for layer by layer pre-training of GBRBM parameters is given in \textbf{Algorithm \ref{alg1}}. 
\begin{algorithm}\label{alg1}
	\SetAlgoLined
	\SetKwInOut{Input}{Input}
	\Input{epoch number, $v$, $K$, GBRBM number, $L$}
	\SetKwInOut{Output}{Output}
	\Output{$\nabla_o w_{ij}, \nabla_o c_j$}
	\BlankLine
	Initialize $\nabla w_{ij}, \nabla b_i, \nabla c_j$ randomly, $\mathbf{v}_{dm}^1\leftarrow \mathbf{v}$\\
	\While{for each epoch}{
	\While{$ m \leq M$}
	{Calculate $h_{dm}^m$ from (\ref{eq5})\\
	 Calculate $v_{dm}^m$ from (\ref{eq6})\\
	 Update the value of $\nabla w_{ij}, \nabla b_i, \nabla c_j$ using (\ref{eqn:parameter2})
	}
	Repeat until the convergence is met
	}
	\caption{Pre-training the GBRBM-blocks (unsupervised learning)}
\end{algorithm}

The autoencoder is used here to reconstruct the input data points that do not have a class label, and the output of the autoencoder is defined as follows:
\begin{eqnarray} 
	e(v) = f \left(c_j + \sum_{i}^{} w_{ij} \frac{v_i}{\sigma_i^2} \right).
\end{eqnarray}

Whereas, the output of the decoder of the hidden layer can be obtained via:
\begin{eqnarray}\label{eq27}  
	r(v) = f \left(c_j^\prime + \sum_{i}^{} \sigma_i^2 w_{ij} e(v) \right).
\end{eqnarray}

Afterwards, the mean-square-error (MSE) cost function is used in the fine-grained phase to optimize the proposed algorithm through the backpropagation algorithm:
\begin{eqnarray}
	Error(D) = \frac{1}{N} \sum_{i=1}^{N}  (r(v)-v)^2
\end{eqnarray}
where $N$ represents the data inputs. Subsequently, a SoftMax classifier is utilized to determine each individual class to which the input belongs. Subsequently, cross-entropy loss is used as a loss function. Hence, \textbf{Algorithm \ref{alg2}} summarizes the aforementioned unsupervised training steps of the autoencoder. 

\begin{algorithm}\label{alg2}
	\SetAlgoLined
	\SetKwInOut{Input}{Input}
	\Input{epoch number, $v$, $K$, GBRBM number, $L$}
	\SetKwInOut{Output}{Output}
	\Output{$\nabla_o w_{ij}, \nabla_o c_j$}
	\BlankLine
	Initialize $\nabla_e w_{ij} \leftarrow \nabla w_{ij} , \nabla_e b_i \leftarrow \nabla b_i,  \nabla_e c_j \leftarrow \nabla c_j$\\
	\While{for each epoch}{
		\While{$ m \leq M$}
		{Estimate the decoded output using (\ref{eq27})}
		Estimate the MSE using (\ref{eq12}) \\
		Fine tune  $\nabla_e w_{ij}, \nabla_e b_i,  \nabla_e c_j$ using backpropagation \\
		Repeat until the convergence is met
	}
	\caption{Fine tuning of autoencoder (unsupervised learning)}
\end{algorithm}

To utilize the detection error between the predicted and real labels, we use the following loss function \cite{Cho2013a}:   
\begin{eqnarray}\label{eqn:loss_function}
	L(w_o,c_o) = \frac{1}{N} y_i \log\left( y_i^\prime(w_o,c_o)\right) 
\end{eqnarray}
where $w_o$ and $c_o$ denote the encoder parameters, while $y$ and $y_0$ represent the predicted and real labels. Thus, the detection error between the predicted and real labels can be estimated using (\ref{eqn:loss_function}). Typically, optimizing parameters and creating classification decisions can be carried out by minimizing the loss function using the enhanced gradient algorithm with the backpropagation algorithm. Since both GBRBM and autoencoder are capable of processing real value data, i.e., GBRBM-based autoencoder (GBRBM-AE) is also applicable of real value data processing. To this end, \textbf{Algorithm \ref{alg3}} is developed for the aforementioned supervised fine-tuning parameters.

\begin{algorithm}\label{alg3}
	\SetAlgoLined
	\SetKwInOut{Input}{Input}
	\Input{epoch number, $v$, $K$, GBRBM number, $L$}
	\SetKwInOut{Output}{Output}
	\Output{$\nabla_o w_{ij}, \nabla_o c_j$}
	\BlankLine
	Initialize $\nabla_o w \leftarrow \nabla_e w^m , \nabla_o b \leftarrow \nabla_e b^m,  \nabla_o c \leftarrow \nabla_e c^m$\\
	\While{for each epoch}{
		Using Backpropagation to estimate the fine tuning $\nabla_o c, \nabla_o w$ \\
		Repeat until the convergence is met
	}
	\caption{Fine tuning of the data labels (supervised learning)}
\end{algorithm}
\section{Performance Evaluation} \label{Sec:performance_evaluation} 
In this section, simulation results are provided to evaluate the performance of the proposed CSI prediction scheme for air-ground links in UAV communication systems. 

\subsection{Experiment Setup}
For the considered system model, the following experiment is designed and performed to obtain practical data measurements from a real operating ground-UAV communication system. 
In the experiment setup, the antennas of the BSs are typically located on tall ground antenna masts, building rooftops, or sometimes hills. 
Within this cellular network, the UAV flies in a wide range of altitudes from ground level to around 300 meters, which is experiencing severe and diverse signal attenuation from different obstacles, (e.g. buildings, trees etc.), different atmospheric conditions (e.g. humidity), long distance from the BSs and loss of ground BS antennas main lobe. Furthermore, as the UAV ascends, more LoS communications can be achieved with different BSs, resulting to increased levels of interference and signal quality degradation.  
The used UAV is a quadcopter utilizing a PX4 flight controller/autopilot and a GPS, with a total take-off weight of almost one kilogram, capable of flying to higher altitudes up to 300 meters and transmitting flight data in real time through its telemetry system. The quadcopter has embedded  measurement unit, a mobile handset with an embedded software for measuring the LTE signal parameters like the received signal reference power (RSRP) of the serving and the neighboring LTE cells.

\begin{figure}[!t] 
\centering 
	\includegraphics[width=0.4\textwidth]{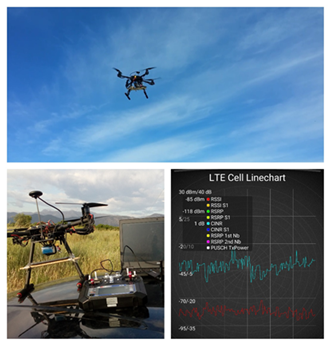}
	\caption{Aerial LTE measurements using a quadcopter.}
	\label{fig:Fig6}
\end{figure}

The mobile handset is connected to the town LTE network using SIM card. In every measurement area, the UAV is flying from the ground level to the altitude of 300 meters and the UAV measurement unit is recording the received signal parameters throughout the flight as shown in Fig. \ref{fig:Fig6}. Combining the data from the UAV GPS for its position and the recorded signal parameters from the onboard measurement device, data sets are created for the development and validation of the proposed DL-based framework.

\subsection{Numerical Results}
The proposed GBRBM-based DNN scheme is implemented and fed with the collected data measurements, i.e. the total number of training vectors is 710, while the number of test vectors is 177, with 201 of unknown variables. 



\begin{figure}[!t] 
\centering 
	\includegraphics[width=0.45\textwidth]{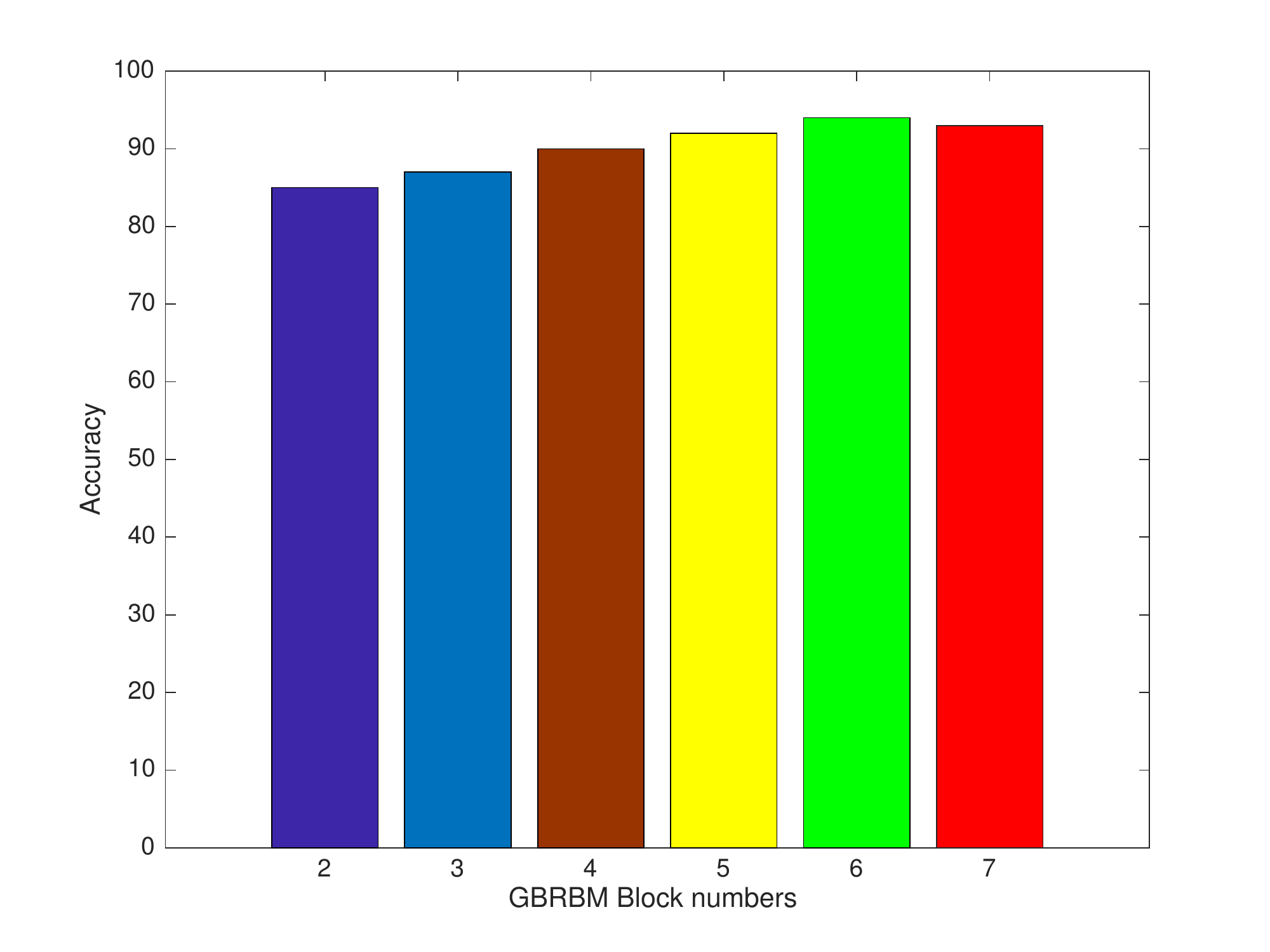}
	\caption{GBRBM-based DNN in the pre-training phase versus the accuracy.}
	\label{fig:Fig7}
\end{figure}

The optimal number of GBRBM-based DNN blocks for the pre-training phase is evaluated in  Fig. \ref{fig:Fig7} by comparing the measurements and the estimated values. 
Different numbers of GBRBM blocks are considered starting from 2 till 7, while the errors between the estimated and measured RSS varies from $-5$ to 15 dBm. It can be readily seen that the best results are obtained when 6 blocks of GBRBM-based DNNs are used in the pre-training phase. The estimation accuracies  are 85.1\%, 87.3\%, 90.1\%, 92.8\%, 94.1\%, and 93.7\%, respectively.

Next, the prediction error versus the number of epochs is evaluated and presented in  Fig. 4. In the pre-training phase, the difference in the RSS values between the estimated measure and the real measure decreases with the increasing of the number of epochs. Additionally, it can be clearly seen that the difference error is stable around the \nth{500} epoch. Hence, the epoch number of the pre-training phase is set to 250 and the epoch number of the training phase is set to 500 epochs.  
After the GBRBM blocks and epoch number are determined, the neuron number must be set for every layer. However, finding the optimal number of neurons is a nontrivial task because the tuning assortment of the neuron number is arbitrary in every layer. Thus, we empirically set the number neurons and  run experiments to maximize the neuron amount of the sixth layer depending on the operation of the GBRBM-based DNN. 


\begin{figure}[!t] 
\centering 
	\includegraphics[width=0.52\textwidth]{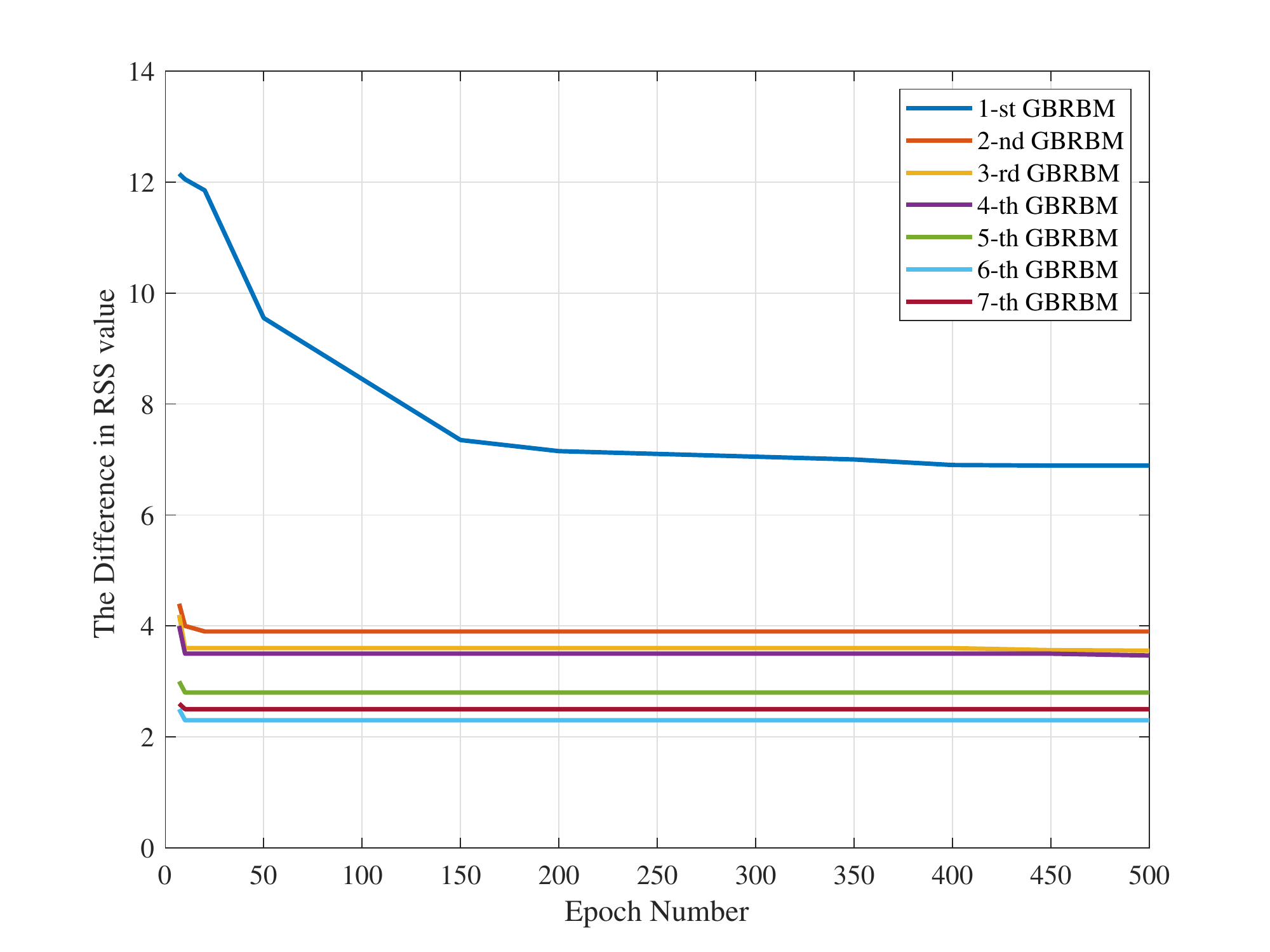}
	\caption{The average difference between the estimated and real RSS values in the pre-training phase.}
	\label{fig:Fig9}
\end{figure}

\begin{figure}[!t]
 \centering 
 	\includegraphics[width=0.5\textwidth]{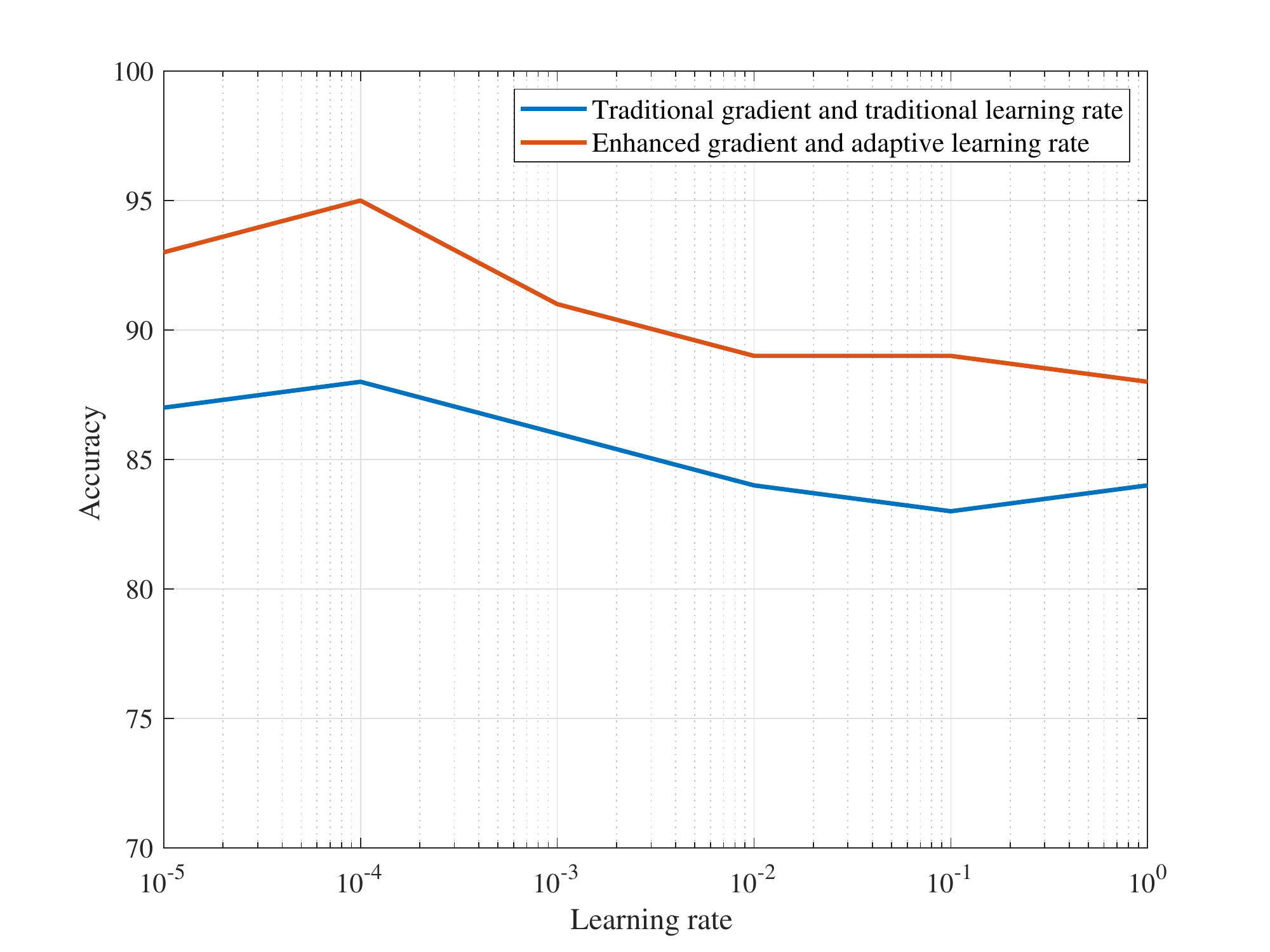}
 	\caption{Accuracy versus adaptive learning rate.}
 	\label{fig:Fi10}
 \end{figure}

The learning rate is similar to the step size of the gradient descent process, namely if it is set too big or too small, the precision will be significantly affected. In particular, if the learning rate is too small, not only does the training period grow but also a local optimal solution is likely to be trapped. To examine the proposed adaptive learning rate approach, we have trained the RBMs of the hidden neurons with the traditional gradient and the same five values (1, 0.1, 0.01, 0.001, 0.0001) to initialize the learning rate. The adaptive learning rate performance during learning is presented in  Fig. \ref{fig:Fi10}. The process can find suitable learning rate values when the enhanced gradient is used. Specifically, one can find 6 GBRBM blocks for the pre-training phase and 5 network layers for the training in the autoencoder. The neuron numbers for hidden layers of multi-block GBRBM are 64, 56, 48, 32, and 16, respectively. 
The learning speeds of the two phases are equally set to 0.001.
 
 \begin{figure}[!t]
	\includegraphics[width=0.5\textwidth]{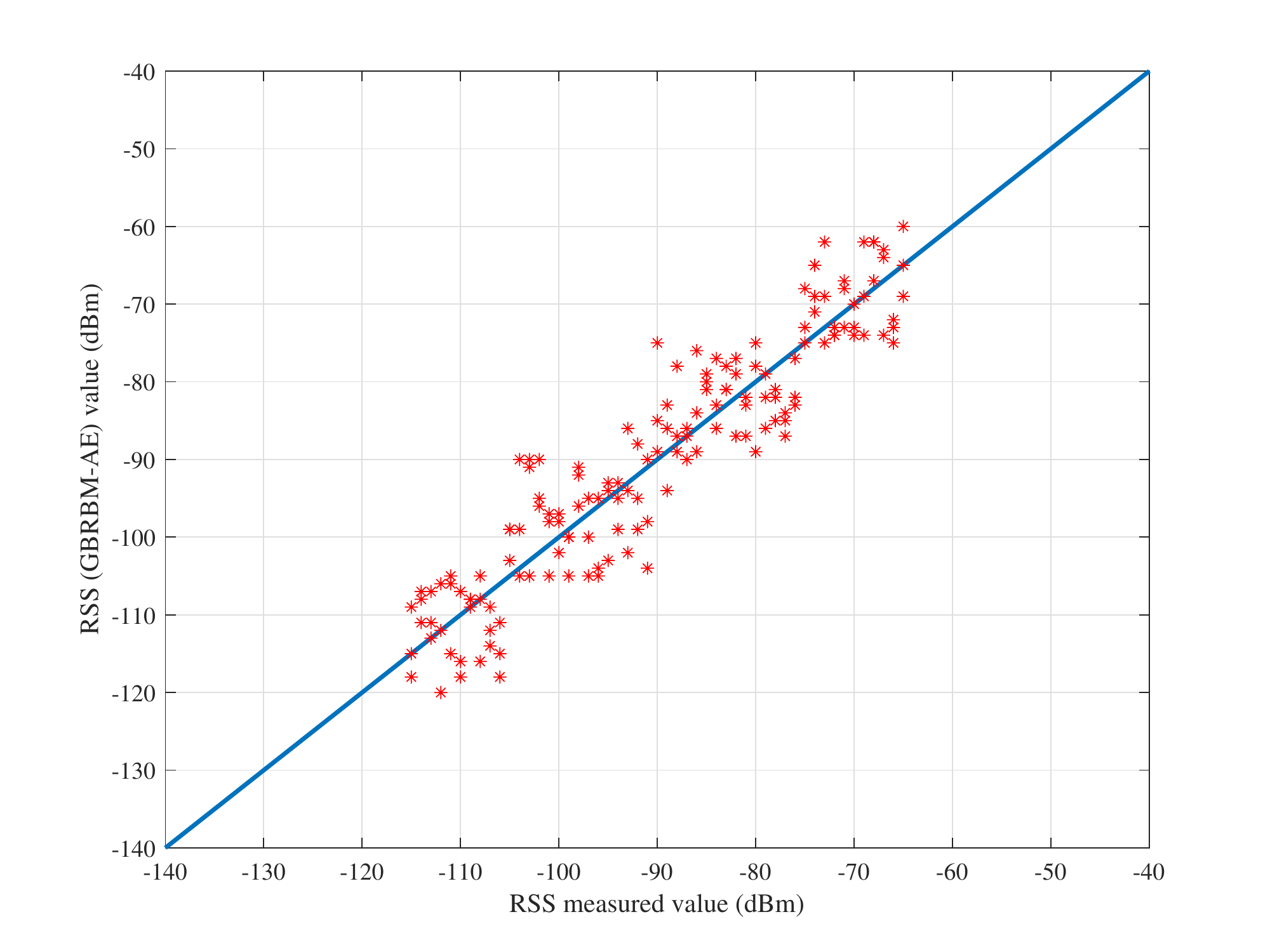}
	\caption{Difference in the RSS values of estimated (GBRBM-AE) and real RSS values in dBm.}
	\label{fig:Fi15}
\end{figure}

To reveal the efficiency of the proposed GBRBM-based autoencoder (GBRBM-AE), the simulation parameters are set similar to the training and pre-training algorithms in order to compare the results for 50 independent trails. The obtained results are shown in Fig. \ref{fig:Fi15}, where the red dots are the measurement values and the blue line is the GBRBM-AE outputs. It can be easily noticing that the predicted values are adequately close to the measurement values. Accordingly, these simulation results show that the proposed algorithm obtains accurate RSS values that can be used for CSI acquisition in various UAV scenarios.

\section{Conclusions}\label{sec:conclusions}
In this paper, a DL-based framework is developed to estimate the channel characteristics of the air-ground links using an UAV flying within a range of altitudes and communicating with a terrestrial network. This framework aims at mitigating the negative impacts of the time-varying environment and differential transmission delay by employing a GBRBM integrated with an autoencoder-based DNN. Although the superiority of RBMs in exploring the latent features in an unsupervised manner, its training is challenging as the stochastic gradient tends to high variance and diverging behavior, and the learning rate has to be manually set according to the RBMs trained structure. 
To circumvent these issues, a novel algorithm is proposed uses adaptive learning rate alongside with an enhanced gradient. The enhanced gradient, contrary to the traditional gradient descent, is used to expedite the learning of the hidden neurons. Finally, the validity of the proposed framework is corroborated by using real UAV signal measurements, and the experimental results have verified the accuracy of our method in learning the UAV channel model in a dynamic propagation environment.

\linespread{1}

\bibliographystyle{IEEEtran}
\bibliography{IEEEabrv,References}

\begin{thebibliography}{10}
\providecommand{\url}[1]{#1}
\csname url@samestyle\endcsname
\providecommand{\newblock}{\relax}
\providecommand{\bibinfo}[2]{#2}
\providecommand{\BIBentrySTDinterwordspacing}{\spaceskip=0pt\relax}
\providecommand{\BIBentryALTinterwordstretchfactor}{4}
\providecommand{\BIBentryALTinterwordspacing}{\spaceskip=\fontdimen2\font plus
\BIBentryALTinterwordstretchfactor\fontdimen3\font minus
  \fontdimen4\font\relax}
\providecommand{\BIBforeignlanguage}[2]{{%
\expandafter\ifx\csname l@#1\endcsname\relax
\typeout{** WARNING: IEEEtran.bst: No hyphenation pattern has been}%
\typeout{** loaded for the language `#1'. Using the pattern for}%
\typeout{** the default language instead.}%
\else
\language=\csname l@#1\endcsname
\fi
#2}}
\providecommand{\BIBdecl}{\relax}
\BIBdecl

\bibitem{Zhao2019}
N.~Zhao, W.~Lu, M.~Sheng, Y.~Chen, J.~Tang, F.~R. Yu, and K.-K. Wong,
  ``{UAV}-assisted emergency networks in disasters,'' \emph{{IEEE} Wireless
  Commun.}, vol.~26, no.~1, pp. 45--51, 2019.

\bibitem{Alladi2020}
T.~Alladi, Naren, G.~Bansal, V.~Chamola, and M.~Guizani, ``{SecAuthUAV}: A
  novel authentication scheme for {UAV}-ground station and uav-uav
  communication,'' \emph{{IEEE} Trans. Veh. Technol.}, vol.~69, no.~12, pp.
  15\,068--15\,077, 2020.

\bibitem{Kisseleff2020}
S.~Kisseleff, W.~A. Martins, H.~Al-Hraishawi, S.~Chatzinotas, and B.~Ottersten,
  ``Reconfigurable intelligent surfaces for smart cities: Research challenges
  and opportunities,'' \emph{IEEE Open Journal of the Communications Society},
  vol.~1, pp. 1781--1797, 2020.

\bibitem{Abdalla2021}
A.~S. Abdalla and V.~Marojevic, ``Communications standards for unmanned
  aircraft systems: The 3gpp perspective and research drivers,'' \emph{IEEE
  Commun. Stds. Mag.}, vol.~5, no.~1, pp. 70--77, 2021.

\bibitem{NGSO_surevy}
H.~Al-Hraishawi, H.~Chougrani, S.~Kisseleff, E.~Lagunas, and S.~Chatzinotas,
  ``A survey on non-geostationary satellite systems: The communication
  perspective,'' \emph{{IEEE} Commun. Surveys Tuts.}, 2022.

\bibitem{Zeng2019}
Y.~Zeng, Q.~Wu, and R.~Zhang, ``Accessing from the sky: A tutorial on {UAV}
  communications for {5G} and beyond,'' \emph{Proc. of the IEEE}, vol. 107,
  no.~12, pp. 2327--2375, 2019.

\bibitem{Khawaja2019}
W.~Khawaja, I.~Guvenc, D.~W. Matolak, U.-C. Fiebig, and N.~Schneckenburger, ``A
  survey of air-to-ground propagation channel modeling for unmanned aerial
  vehicles,'' \emph{{IEEE} Commun. Surveys Tuts.}, vol.~21, no.~3, pp.
  2361--2391, 2019.

\bibitem{Sun2021}
L.~Sun, Y.~Wang, A.~L. Swindlehurst, and X.~Tang,
  ``Generative-adversarial-network enabled signal detection for communication
  systems with unknown channel models,'' \emph{{IEEE} J. Sel. Areas Commun.},
  vol.~39, no.~1, pp. 47--60, 2021.

\bibitem{Ma2020}
X.~Ma and Z.~Gao, ``Data-driven deep learning to design pilot and channel
  estimator for massive {MIMO},'' \emph{{IEEE} Trans. Veh. Technol.}, vol.~69,
  no.~5, pp. 5677--5682, 2020.

\bibitem{Choo2018}
S.~Choo and H.~Lee, ``Learning framework of multimodal gaussian–bernoulli rbm
  handling real-value input data,'' \emph{Neurocomputing}, vol. 275, pp.
  1813--1822, 2018.

\bibitem{Koller2009}
D.~Koller and N.~Friedman, \emph{Probabilistic Graphical Models: Principles and
  Techniques - Adaptive Computation and Machine Learning}.\hskip 1em plus 0.5em
  minus 0.4em\relax The MIT Press, 2009.

\bibitem{Carreira2005}
M.~A. Carreira-Perpi{\~n}\'an and G.~Hinton, ``On contrastive divergence
  learning,'' in \emph{The 10th Int. Workshop on Artificial Intelligence and
  Statistics}, ser. Proceedings of Machine Learning Research, vol.~R5, Jan.
  2005, pp. 33--40.

\bibitem{Cho2013}
K.~Cho, T.~Raiko, and A.~Ilin, ``Enhanced gradient for training restricted
  boltzmann machines,'' \emph{Neural Computation}, vol.~25, no.~3, pp.
  805--831, 2013.

\bibitem{Cho2013a}
K.~H. Cho, T.~Raiko, and A.~Ilin, ``Gaussian-bernoulli deep boltzmann
  machine,'' in \emph{International Joint Conference on Neural Networks
  (IJCNN)}, 2013, pp. 1--7.

\end{thebibliography}
	
\end{document}